# Ballistic dispersive shock waves in optical fibers

**J. Nuño[1,2], C. Finot[1], G. Millot[1], M. Erkintalo[3] and J. Fatome[1,***

[1]Laboratoire Interdisciplinaire Carnot de Bourgogne, UMR 6303 CNRS - Université Bourgogne-Franche-Comté, 9 av. A. Savary, Dijon, France

[2] Now at Departamento de Electrónica, Universidad de Alcalá, 28805 Alcalá de Henares (Madrid), Spain

[3]Dodd-Walls Centre and Department of Physics, The University of Auckland, Private Bag 92019, Auckland 1142, New Zealand

*Corresponding author: julien.fatome@u-bourgogne.fr

**We experimentally and numerically report on the formation of ballistic dispersive shock waves that mimic superfluid behavior in optical fibers. The shock wave is triggered by a strong nonlinear phase shift induced by a high power pump pulse on a cross-polarized continuous-wave probe field, and it exhibits a central void surrounded by two oscillating fronts moving away from each other with opposite velocities. The impact of disorder in the continuous-wave background as well as the difference of group-velocities between the two orthogonally polarized waves is also investigated. Our experimental observations are in very good agreement with theoretical and numerical predictions derived from the Manakov model.**

## 1. INTRODUCTION

The formation of dispersive shock waves (DSWs) is a fundamental mechanism encountered in many fields of science, such as hydrodynamics, geophysics, supersonics, atmospherics, socioeconomics, chemistry, and nonlinear optics [1-6]. One of the most fascinating manifestations of DSWs in nature is the appearance of the popular Mascaret wave which can be generated in specific river estuaries and for which a counter-flow between tide and current is formed, generating large undular tidal bores travelling up-stream and traditionally well-appreciated by the surfers community [1, 2]. Another spectacular manifestation of DSWs in atmospheric air-flow relies on the emergence of Morning Glory roll clouds and mountain waves [3]. In general, DSW phenomena occur in conservative (or weakly dissipative) system, and they rely on two fundamental ingredients: nonlinearity and wave dispersion. The most common situation exhibits the dispersive regularization of a jump gradient catastrophe, giving rise to the expansion of a non-stationary oscillating fan structures connecting both upper and lower initial non-oscillating states [5, 6]. Because of their universal nature, DSW prototypes have become the subject of intense research in physics, with groundbreaking experiments realized in diverse areas such as in fluid dynamics, condensates, and nonlinear optics [6-17].

In a Kerr nonlinear optical medium, under the assumption of a weakly self-defocusing dispersive regime, the photon flow acts as an ideal fluid or gas and can thus be modeled by a shallow water wave equation for which any incident modulation experiences a strong steepening reshaping, thus leading to the formation of a gradient catastrophe and subsequently DSWs [18-20]. It is important to notice that this latter is a universal law, not restricted only to nonlinear optics, in the sense that it occurs whenever the wave velocity becomes function of the wave elevation (here the Kerr effect) and can thus be summarized as a dispersive hydrodynamics problem [6]. However, the difficulty to generate and characterize in laboratory pure DSWs is still a challenging issue. Indeed, since the pioneering experiments reported by Rothenberg in 89 [9], huge efforts have been realized to develop powerful test-bed platforms capable of mimicking fluid-type DSWs with nonlinear optics [10-17]. In most of these previous works, the DSWs is generated through the nonlinear steepening of an intense bright pulse superimposed on a continuous-wave background (cw) propagating in a defocusing (normal dispersion) nonlinear Kerr medium [16, 21]. Subsequently, high frequency components are generated on the edges of the pulse and interfere with the cw background so as to trigger the generation of DSWs. However, with such a configuration, a part of the phenomenon is hidden by the pulse itself and more importantly, neither ballistic shock waves (brutal generation of temporal gaps) nor piston problems can be emulated. Consequently, in order to generate pure depression DSWs and fully mimic superfluid dynamics, one has to dissociate the DSW formation from the high intensity bright pulse in such a way as to imprint the dispersive hydrodynamics phenomenon only on the cw landscape.

To this aim, here we propose to exploit the vectorial dimension offered by optical fibers so as to imprint an expanding gradient catastrophe on the phase profile of a cw landscape through a cross-polarized phase modulation process (XPM). More precisely, as illustrated in Fig. 1, the key idea consists of a vectorial XPM interaction between a weak cw probe that is co-propagating in a normally dispersive optical fiber together with an orthogonally polarized intense short pump pulse. The defocusing regime of the fiber induces a strong steepening of the pulsed signal. This increasing sharpness and temporal expansion of the pulse edges then triggers an expanding catastrophe chirp profile on the orthogonally polarized CW probe through the XPM coupling. Forced by this piston effect, this nonlinear phase gradient catastrophe is then regularized by chromatic dispersion, removing all energy contained in the central region of the probe while generating two repulsive DSWs superimposed onto the initial CW landscape, thus mimicking superfluid DSWs. To the best of our knowledge, our results represent the first experimental observations of vectorial DSWs generated in optical fibers, and more generally, the first observation of what we have chosen to name a ballistic shock wave in nonlinear optics. We have also investigated the impacts of group-velocity mismatch and randomness in the CW landscape to the DSW profile and dynamics, obtaining results in excellent agreement with numerical simulations and theoretical predictions. Our work provides significant insights on DSWs, and demonstrates a convenient platform that allows for the systematic experimental study of DSW physics.

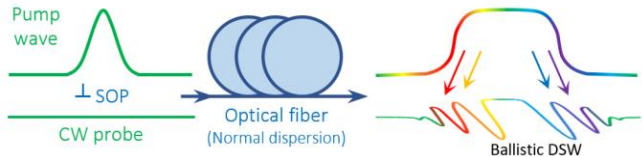

Fig. 1. Schematic of the ballistic dispersive shock wave phenomenon.

## 2. MODELING AND PRINCIPLE OF OPERATION

**A. Modeling**

The system under study exploits the vectorial XPM interaction between a weak CW probe and an orthogonally polarized intense short pulse, both of which propagate in a normally dispersive optical fiber. In the framework of nonlinear fiber optics [22], the evolution of the complex slowly varying amplitudes of the pulsed signal $u$ and the CW probe $v$ are described by the following set of two coupled nonlinear Schrödinger (NLS) equations . In standard optical fibers, random spatial fluctuations of the residual birefringence occur in a length-scale of a few meters. Averaging out the nonlinear contribution over these fast polarization fluctuations in km-long fibers leads to the so-called Manakov model [23, 24]:

$$\begin{cases} i\dfrac{\partial u}{\partial z} + \dfrac{\beta_2}{2}\dfrac{\partial^2 u}{\partial t^2} + \dfrac{8}{9}\gamma\left(|u|^2+|v|^2\right)u + i\dfrac{\alpha}{2}u = 0, \\ i\dfrac{\partial v}{\partial z} + i\delta\dfrac{\partial v}{\partial t} + \dfrac{\beta_2}{2}\dfrac{\partial^2 v}{\partial t^2} + \dfrac{8}{9}\gamma\left(|u|^2+|v|^2\right)v + i\dfrac{\alpha}{2}v = 0 \end{cases} \quad (1)$$

Here, $z$ and $t$ denote the propagation distance coordinate and time in the co-moving frame of the pulse, respectively. $\gamma$ corresponds to the nonlinear Kerr parameter of the fiber, $\beta_2$ is the group velocity dispersion coefficient, and $\alpha$ indicates propagation losses. The corrective factor of 8/9 is applied to the Kerr effect due to the polarization randomness assumed in the Manakov model. The term $\delta = 2\pi\beta_2\delta f$ describes the walk-off between the pulse and the CW landscape when their central frequencies are separated by $\delta f$. Let us stress that no transfer of energy occurs between the two waves during the nonlinear interaction. Indeed, as shown by the fourth term in the second Eqs. (1), both waves are only coupled by a phase term proportional to the temporal intensity profiles of the waves. Accordingly, the energy initially contained in the CW probe (or pulsed pump) landscape remains unchanged (except for linear losses). Note that we neglect higher-order effects such as third-order dispersion, self-steepening and Raman scattering, as they have been found to play a minor role in the present dynamics. This assumption is confirmed below by the good agreement between our experimental observations and numerical modeling.

Equations (1) can be further simplified via additional approximations whose validity we have already confirmed in vectorial applications such as high-repetition sources [25] or optical sampling [26]. As first approximation, we may neglect the consequences of fiber losses $\alpha = 0$. We also consider that the $u$ and $v$ have identical central frequencies so that the walk-off parameter is null $\delta = 0$. Moreover, as the initial peak-power of the pulse $u$ is much larger than the power of the probe $v$, we can neglect the cross- and self-phase modulations involved with the weaker probe signal. With such assumptions, Eq. (1) can be cast into the following coupled reduced equations, with $\gamma' = 8\gamma/9$:

$$\begin{cases} i\dfrac{\partial u}{\partial z} + \dfrac{\beta_2}{2}\dfrac{\partial^2 u}{\partial t^2} + \gamma'|u|^2 u = 0, \\ i\dfrac{\partial v}{\partial z} + \dfrac{\beta_2}{2}\dfrac{\partial^2 v}{\partial t^2} + \gamma'|u|^2 v = 0 \end{cases} \quad (2)$$

Note that in the following, for convenience, we may call the pulse $u$, the pump even if no transfer of energy occurs between $u$ and $v$.

In order to illustrate the formation and evolution of the shock wave in a vectorial configuration, we consider the following set of parameters which corresponds to the experiment described in section III. The pump is a Gaussian-like chirp-free pulse characterized by a temporal half-width at $1/e$ of $t_0 = 41$ ps, a peak power $P_c = 1.5$ W and a central wavelength of 1550 nm. The average power of the orthogonally polarized CW probe is 10-mW. The fiber is a 13-km-long dispersion compensating fiber (DCF) typically used in the field of dispersion managed data transmission. The DCF fiber spool is characterized by a chromatic dispersion $\beta_2 = 166$ ps$^2$/km (D = −130 ps/nm/km) at 1550 nm, an attenuation parameter $\alpha = 0.4$ dB/km and a nonlinear Kerr coefficient $\gamma' = 5.5$ W$^{-1}$.km$^{-1}$. It is important to notice that these experimental parameters place our problem in the weakly dispersive regime of the NLS (at least in the first stage of propagation), for which the ratio between nonlinear and dispersive effects is given by the usual soliton number $N = \sqrt{\gamma' P_c t_0^2/\beta_2} \sim 14$ at maximum pump power [22]. Therefore, our study fully satisfies the analogy between nonlinear optics and shallow water hydrodynamics in the early stage of propagation [18].

## B. Dynamics of the dispersive ballistic shock wave generation

Figure 2 shows typical numerically simulated dynamics of ballistic dispersive shock wave generation. Give our reduced model [Eq. (2)], the pump pulse is here unaffected by the probe and therefore experiences nonlinear dynamics typical for the evolution of a high power pulse in a normally dispersive medium. Specifically, because of the highly nonlinear defocusing regime of propagation, the input pump field $u(t)$ initially experiences strong self-phase modulation evolution induced by the dominant nonlinear Kerr term present in first Eqs. (2) [22], as depicted by the longitudinal evolution of the pump spectrum in Fig. 2(a). Then, due to the interplay with chromatic dispersion, the pump pulse undergoes a temporal expansion (panel b) as well as a strong envelope steepening. Subsequently, the pump is reshaped into broad sharp pulses and parabolic profiles with the formation of a so-called gradient catastrophe, leading to characteristic wave-breaking and shock-wave behavior [9, 16, 27, 28] that occurs at a distance $z_c$ indicated by means of a cyan dashed-line in Fig. 2(a):

$$z_c \simeq \frac{1.61\, t_0}{\sqrt{\beta_2\, \gamma'\, P_c}} \qquad (3)$$

Since the optical spectra of each waves have become much broader, chromatic dispersion tends to dominate with subsequent propagation and, in the far-field of propagation, induces a dispersive mapping of the whole process [29, 30]. Based on previous studies dealing with wave-breaking, in strong contrast to a pure nonlinear regime for which the nonlinear phase-shift induced by SPM is proportional to the peak-power and propagation distance [22], here the competition between nonlinear effect and chromatic dispersion leads to a parabolic-like spectrum [29] with a maximal spectral broadening which can be evaluated by the following expression [21, 30] and indicated with open circles in Fig. 2(a):

$$\Delta\omega_{max} = 2^{3/4} \sqrt{\frac{\gamma'\, P_c}{\beta_2}} = 2^{3/4}\, \frac{N}{t_0} \qquad (4)$$

The temporal expansion of this non-linear dispersive similariton pump wave $\Delta t_{max}$ follows the dispersive mapping induced by the frequency dependence of the wave velocities due to chromatic dispersion. Therefore, as exploited in the dispersive Fourier transform technique [31], the temporal intensity profile of the pump becomes a scaled replica of the parabolic spectral intensity profile [29, 32], while the longitudinal evolution of $\Delta t_{max}$ is given by:

$$\Delta t_{max} = 2^{3/4}\, z\, \sqrt{\gamma'\, P_c\, \beta_2} \simeq 2.71\, z/z_c \qquad (5)$$

This last relation fits very well with the temporal expansion of the pump pulse observed in numerical simulations, as shown in Fig. 2(b).

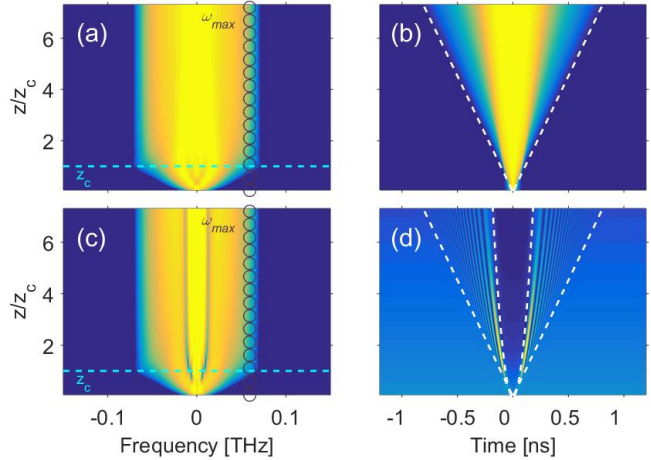

Fig. 2. (a) and (c): Longitudinal evolution of the spectrum of the pump and probe waves, respectively. The cyan dashed-line indicates the distance $z_c$ predicted by Eq. (3) while the purple open circles highlight $\omega_{max}$ predicted by Eq. (4). (b) and (d): Longitudinal evolution of the intensity profiles of the pump and probe waves, respectively. Data values of the cw probe are normalized by the input level of the CW landscape while the pump profile is normalized by the peak power. Analytical predictions of the temporal expansion provided by Eqs. (5) and (6) are plotted with white dashed lines. Results obtained for a pump peak power of 1.5 W and a cw background of 10 mW, $z_c$ = 1.785 km.

In parallel with the evolution of the pump wave, the XPM interaction causes the CW probe beam *v(z,t)* to progressively acquire a nonlinear phase shift $\phi_{nl}(z,t)$ proportional to the intensity profile of the pump $|u(z,t)|^2$ [33]. This latter implies the appearance of an instantaneous frequency deviation (chirp) within the CW landscape $\delta\omega(z,t) = -\partial_t \phi_{nl}(z,t) = -\gamma z \, \partial_t |u(z,t)|^2$ characterized by a redshift on the leading part and a blueshift throughout the trailing edge, leading to a large spectral broadening (Fig. 2(c)). Since in a normally dispersive medium, blue light travels slower than red, such a chirp is turned into an instantaneous change of velocity $V$ around the central point of the probe/pump pulse $\delta V(z,t) \propto -\delta\omega(z,t)$ and directed along outward directions. As the pump pulse expands and develops steeper and steeper edges, a piston effect is then imposed into the phase profile of the probe, which triggered by chromatic dispersion leads to the creation of two fronts of opposite velocities and to the removal of energy from the central region (see Fig. 2d). Beyond this point, the chromatic dispersion becomes important near the vertical fronts, causing the onset of fast non-stationary oscillations surrounding a temporal gap, thus creating the characteristic imprint of a ballistic dispersive shock wave.

Quite importantly, given our configuration and in absence of SPM of the probe, the nonlinear chirp experienced by the probe is identical to the SPM undergone by the pump. Therefore, as shown in Figs. 2 (c) & (d), Eqs. (4) & (5) also apply to the probe in order to predict the temporal and spectral spreading. Following a similar procedure, the rarefaction area $\Delta t_{min}$, depicted with a white dashed-line in Fig. 2 (d), can also be predicted from the peak-power reduction of the pump along the propagation distance calculated from its maximal temporal broadening:

$$\Delta t_{min} = 2^{3/4} \; z \; \sqrt{\gamma' \, P_{min} \, \beta_2} \qquad (6)$$

With

$$P_{min} = P_c \frac{t_0}{\Delta t_{max}} \qquad (7)$$

The piston effect that induces the ballistic DSWs is even more striking when illustrated by means of a spectro-temporal approach. To this end, we have computed the spectrogram (i.e. the time-frequency representation) [34, 35] of the waves at different stages of propagation [28]. Results are summarized in Fig. 3. In the first stage of propagation, the pump experiences both temporal and spectral expansion and the interplay between dispersion and nonlinearity progressively leads to the development of a close-to-linear chirp. When $z \gg z_c$, the spectral extent does not evolve anymore and the pump pulse behaves like a spectron, characterized by a close-to-linear chirp profile with a slope given by $1/(\beta_2 z)$ [29].

The dynamics of the cw probe is more complex. In the first stage of propagation, the cross-phase modulation induced by the pump leads to the generation of new frequency components around the CW: components at lower frequency (redshift) are generated in the region corresponding to the leading part of the pump whereas components with higher frequencies (blueshift) are generated in the region corresponding to the trailing edge of the pump. As the pump pulse expands and develops steeper and steeper edges, the normal dispersion comes into play and the piston effect acts in full strength pushing the various components far from the center. This phenomenon creates a depletion of energy in the center, a temporal gap which continuously expands over propagation. On both sides of this rarefaction area, a dispersive shock wave emerges from the fast oscillations occurring between the remaining continuous wave background and new frequencies [16]. This rarefaction area obtained in a single fiber with distributed linear and nonlinear effects can to some extent be compared with the temporal cloaking process relying on two stages: an XPM-induced broadening induced in a lumped highly-nonlinear waveguide followed by purely dispersive propagation [36].

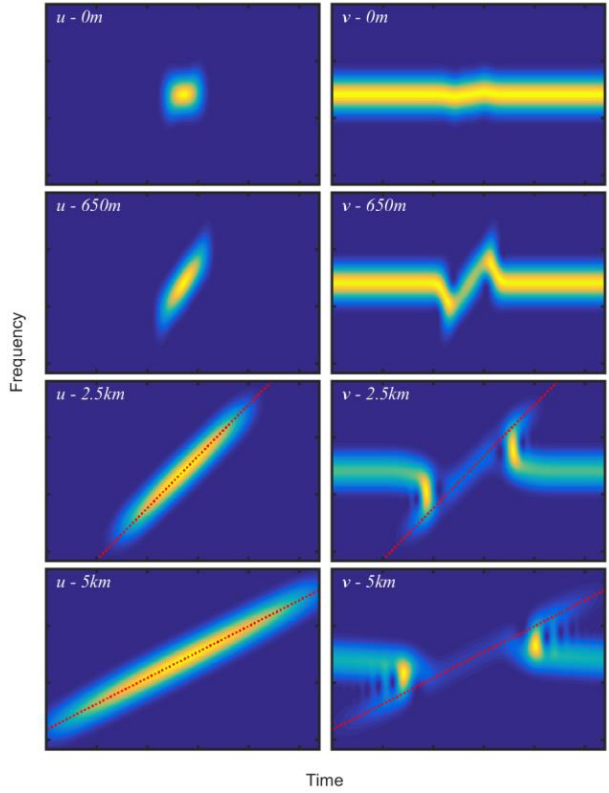

Fig. 3. Evolution of the spectrograms at different propagation distances for the pump (left column) and probe waves (right column), respectively. The red dashed lines represent the far field chirp with a slope of $1/(\beta_2 z)$. The spectrograms are plotted on a linear scale and extend from -300 ps to 300 ps and from -80 GHz to 80 GHz. Results obtained for a pump peak power of 1.5 W.

Contrary to previous works investigating the development of fan structures induced by the nonlinear evolution of a phase-modulated continuous waves [15, 20], many differences are worth stressing in our two-component system. First, in previous works dealing with scalar propagation, the phase modulation was the initial fully-controlled condition of the continuous wave whereas here it is continuously induced by the propagating beam in a non-trivial manner. Second, the self-phase modulation of the flaticon was a key ingredient of the spectro-temporal dynamics, while in the present work, SPM of the seed signal can be neglected and the nonlinear phase modulation is induced not by the wave itself but by the orthogonally propagating pulse. The evolution of the shock wave is therefore intrinsically highly coupled to the evolution of the pump wave. Similar differences hold when comparing our configuration with the recently published pattern [37] resulting from the nonlinear propagation of a dark perturbation in a nonlinearly focusing medium where, in this case, modulation instability plays a crucial role. Finally, because the present ballistic DSW develops onto a continuous wave landscape, the resulting pattern appears very different from the features numerically described in XPM-induced optical wave-breaking [38].

## 3. EXPERIMENTAL VALIDATION

### A. Experimental setup

To experimentally generate and observe ballistic DSWs, we have implemented the setup depicted in Fig. 4. Note that this test-bed platform relies exclusively on commercial devices widely used in the telecommunication industry. A train of 68-ps pulses at a repetition rate of 312.5 MHz is first generated at 1550 nm by means of a CW laser modulated thanks to a 14-Gbit/s pulse-pattern generator (PPG). Two intensity modulators (IM1 & IM2) are cascaded in order to reach an extinction ratio higher than 40 dB and to prevent any spurious interference between the pump and the residual background [16]. This pulse train is then amplified by means of a 33-dBm Erbium doped fiber amplifier (EDFA) and used as pump signal whose power can be continuously tuned thanks to a programmable variable optical attenuator.

The probe wave consists of a 10-mW CW landscape which can be a simple replica of the initial 1550-nm signal to ensure a perfect velocity matching (null walk-off configuration, $\delta = 0$ in Eqs.(1)) or a distinct CW generated from a second external cavity laser so as to induce a variable velocity mismatch between the pump and probe signals. Furthermore, to study the influence of the randomness of the medium, here provided by the coherence properties of the CW landscape, (see section V), the CW can be replaced by a partially coherent wave generated from a spontaneous noise emission source (ASE) followed by a tunable optical bandpass filter and a polarizer.

Pump and probe signals are then orthogonally polarized by means of two polarization controllers (PC), combined in a first polarization beam splitter (PBS), and injected into a 13-km long normally dispersive fiber (DCF) whose parameters were listed in the above modeling section. At the output of the system, the pump and CW probe signals are polarization demultiplexed thanks to a second PBS and characterized both in the temporal and spectral domains with a 70-GHz electrical sampling oscilloscope and an optical spectrum analyzer (OSA). Note that before temporal detection, a tunable optical bandpass filter may be inserted in such a way to carry out a spectral-temporal analysis (see Fig. 9). Finally, we would like to stress here that the all-fibered nature of the setup ensures its stability over several hours of operation.

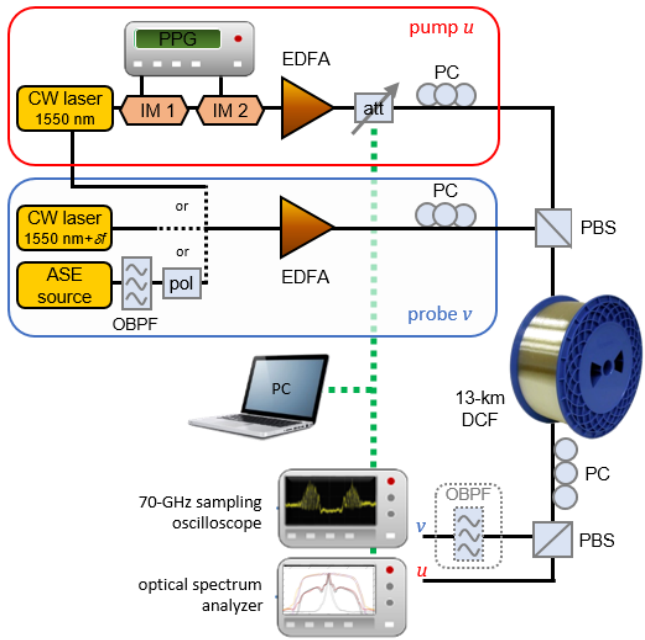

Fig. 4. Experimental setup. CW : Continuous Wave, ASE : Amplified Spontaneous Emission, PPG : Pulse Pattern Generator, IM : Intensity Modulator, EDFA : Erbium Doped Fiber Amplifier, Att : Optical Attenuator, PC : Polarization Controllers, PBS : Polarization beam splitter/combiner, DCF : Dispersion Compensating Fiber, OBPF : Optical BandPass Filter, Pol : Polarizer.

**B. Experimental temporal and spectral profiles**

Figure 5 displays the experimental mapping of the ballistic DSW generated on the CW landscape as a function of the pump peak-power. Here a single initial CW laser is used for both waves (null walk-off configuration). The power variation is directly performed with a remote programmable attenuator, thus ensuring an efficient stability of the setup, especially on polarization states. Moreover, the full automatization of our experimental platform enables us to accumulate up to 150 measurements in a very short time scale, providing a perfect stability during the whole shock process.

Note that this procedure partly mimics the recording of the longitudinal evolution of the fields and avoids the cutback method which is unpractical and fully destructive. In Fig. 5, we can clearly observe the piston effect caused by the pump induced XPM on the probe signal, and in particular, the depletion of the central part caused by the generation of two fronts of opposite velocities, in good agreement with our expectations described above. Indeed, the shock wave formation is accompanied by the appearance of a gap in the temporal domain surrounded by two repulsive oscillating fronts resting on the continuous wave background, confirming the analogy with superfluid or ballistic DSWs.

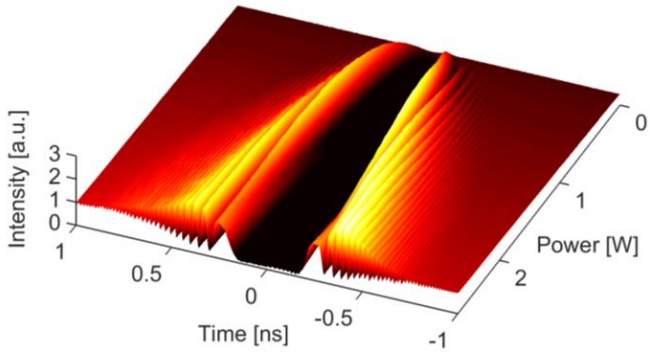

Fig. 5. False-color plot showing the experimental evolution of the output intensity profile of the probe as a function of input pump peak power. Data values are normalized with the initial background equal to 1 and mapped to colors linearly.

The development of the shock fan is even more impressive when visualizing the evolution of the CW landscape in two dimensions. Figures 6 (a&b) depict the experimental evolution of respectively the shock fan and pump pulse as a function of the injected pump power. The experimental results (panels a&b) are in very good agreement with corresponding numerical simulation results (panels c&d). These simulations were obtained using Eqs. (1) with experimental parameters.

We can clearly see the appearance of an expanding temporal gap surrounded by fast non-stationary oscillations that fill the characteristic shock fans. We can also notice the wide temporal broadening of the initial pump pulse which coincides with the expansion of the shock. The theoretical predictions of the DSWs expansion based on Eqs. (5)-(6) are also reported in Fig. 6 (white dashed lines), and they are found to accurately match the temporal broadening of both waves according to the input pump power.

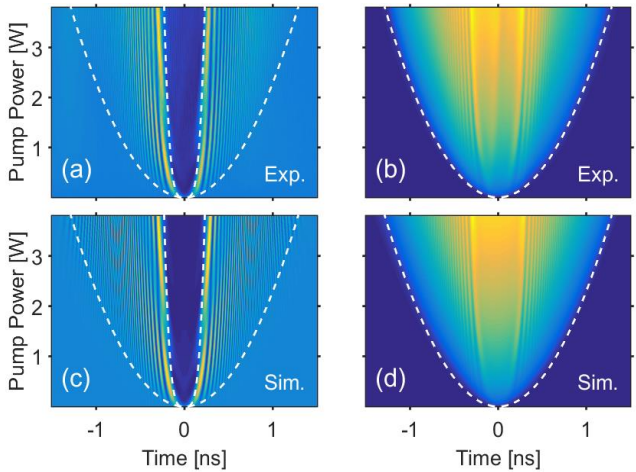

To go further into the analysis, we report in Fig. 7 the experimental recording of the input and output properties of the pump and probe waves for an injected peak-power of 1.5 W. In panel (a), we can first notice the large temporal broadening of the pump pulse: the pulse width increases from 68 ps at half-maximum (blue solid line) to more than 1.4 ns (yellow line). Moreover, the output pulse shape can be well approximated by a parabolic profile of width $\Delta t_{max}$ [Eq. (5)], depicted in Fig. 7(a) with circles, thus confirming the dispersion mapping of the whole process in long propagation distances [30]. We can also notice the presence of small oscillations on the edges of the output pump profile when compared to the case without CW probe (purple line). We attribute the origin of these oscillations to the nonlinear contribution of the ballistic DSW on the pump signal through XPM (neglected in first approximation in Eqs. (2)). Indeed, for large propagation distances, due to the ns broadening experienced by the pump, its peak power has significantly decreased, becoming comparable to the probe peak power. Regarding the probe output intensity profile, displayed in red in Fig. 7(b), the piston effect induced by the pump can be readily observed. The nonlinear phase shifts induced by the expansion and steepening of the pulse edges create two moving fronts starting from the center and propagating in outwards directions.

The nonlinear reshaping of the pump is also accompanied by a strong spectral broadening, shown in Fig. 7(c) in yellow, characterized by an output total spectral expansion above 150 GHz, close to the theoretical value $2\Delta\omega_{max}$ indicated by means of blue dashed lines. Note that the output spectrum of the probe signal in red (initially CW, gray) is also widely broadened with a spectral extension rather similar, underlining the XPM coupling between the ballistic shock wave and the pump. Numerical simulations (in black for the probe and in purple for the pump) again show good agreement with experimental observations. Note that, while the temporal and spectral intensity profiles of the pump are directly linked by the dispersive Fourier transform [31], this is not the case for the probe because of the presence of the continuous background.

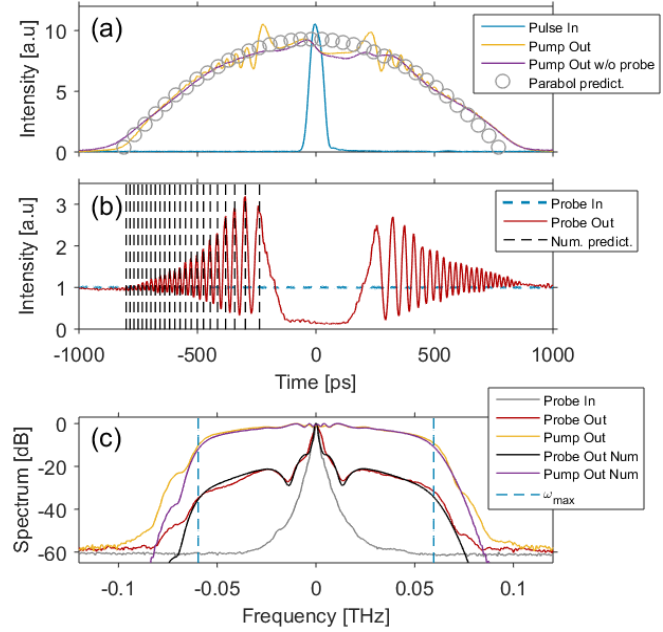

Fig. 7. Experimental results obtained for a pump power of 1.5 W. (a) Temporal profile of the pump wave. The input pulse (blue solid-line) is compared with the output profile in presence (yellow) and without (purple) the cw probe. (Circles) Comparison with a parabolic shape of width

*Δt$_{max}$*. (b) Temporal profile of the probe wave in input (blue dashed line) and output (red solid line). Vertical black dashed lines indicate the periods of oscillations predicted by numerical simulations based on Eqs. (1). Note that in panels (a) and (b), data values are normalized with respect to the cw landscape except for the input pulse profile, here arbitrary normalized for convenience and clarity. (c) Corresponding spectral profiles.

## C. Details of the oscillatory pattern

The ballistic shock clearly develops into a wide open temporal gap surrounded by two expanding strongly modulated wave trains of opposite velocities that link the high humps to the continuous landscape boundaries. It is important to stress that, thanks to our 70-GHz bandwidth detector and the length of the DCF that we have carefully chosen, these oscillating traveling waves are fully resolved. The absence of any major asymmetry, typically provided by third-order dispersion, self-steepening or Raman scattering, also confirms the various assumptions made in the model. However, the slight asymmetry observed in the DSW is attributed to the small asymmetry of the input pulse. Note finally the excellent agreement obtained for the periods of oscillations between our measurements and the ones predicted by numerical simulations based on Eqs. (1), indicated in Fig. 7(b) by means of vertical black dashed-lines. The different periods and temporal positions of these oscillations are also fully consistent with the dispersion-induced spectro-temporal mapping of the shock. Indeed, since based on an interference pattern between the CW components and highly chirp edges, the temporal oscillations of the shock fan can be well approximated as $1 + 2\cos(t^2/2\beta_2 L)$. In order to further highlight this point, we have reported in Fig. 8 the evolution of the pattern frequencies and corresponding temporal positions of these oscillations along the shock fan. The experimental results (here in green stars) match very well these predictions (orange solid line). Moreover, we can also notice the good agreement obtained through numerical simulations for the shock (circles) as well as the corresponding spectro-temporal evolution of the pump (in yellow squares), thus confirming the genuine dispersive mapping of these vectorial interactions.

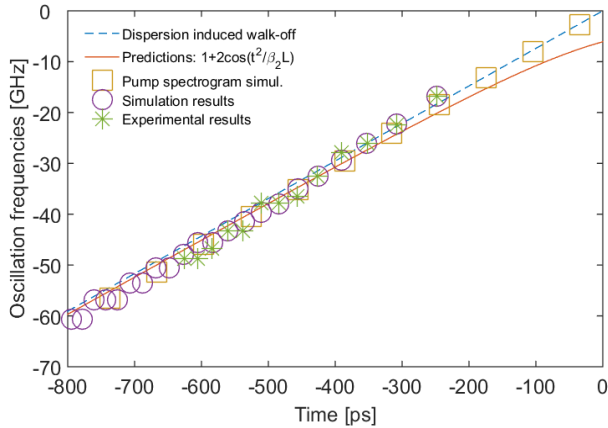

Fig. 8. Temporal mapping of the shock oscillation frequencies. The experimental results (green stars) are compared with numerical simulations (in purple circles) and walk-off induced by chromatic dispersion (blue dashed line). Theoretical predictions are included with an orange solid line while the pump spectrogram, obtained from numerical simulations, is also highlighted by means of yellow squares.

Implementing a frequency-offset optical bandpass filter before the detection enables us to get access to the temporal localization of the components having different frequencies. This method is particularly relevant here especially to detect the expected frequency shift on the edge of the shock, as predicted in Fig. 3. To this aim, Figs. 9(a&b) display the resulting spectrum and corresponding intensity profile recorded after a 1-nm offset filtering. For comparison, the pump spectrum is also depicted in Fig. 9(a) (grey curve).

From these figures, we can observe that filtering induces a striking asymmetry in the temporal intensity profile. Indeed, by filtering the output spectrum of the probe, we are able to annihilate half of the shock. More precisely, after cancelling the higher frequencies (red curve), the temporal fan recorded for $t > 0$ vanishes. In contrast, when higher frequencies and the continuous component are preserved (blue curves), the leading fan disappears. In both case, the central hole still occurs. This behavior is fully consistent with the spectrogram representations obtained numerically in Fig. 3, for which the low- (high-) frequency components are localized in the leading (trailing) fan and that no energy is still present in the central part of the shock.

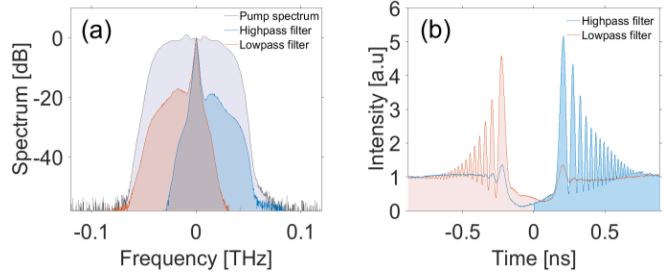

Fig. 9. (a) Output spectrum after optical filtering obtained for a pump power of 1 W. (b) Output temporal intensity profile of the probe signal after spectral filtering. Resulting signal, in red (blue), when only the low (high) frequencies are sliced.

## 4. INFLUENCE OF A VELOCITY MISMATCH

We have also experimentally investigated the impact of the walk-off that may exist due to a frequency offset $\delta f$ between the pump pulse and the continuous landscape. In this situation, the ballistic shock occurs between a fixed pump and a moving probe or vice-versa. For this purpose, instead of using a single CW, we have exploited two separate tunable lasers at different wavelengths which, through chromatic dispersion, travel at different speeds in the fiber.

We can observe in Fig. 10(a) the temporal pattern of the shock observed for a pulse peak-power of 2 W and a probe wave detuned by $\delta f = 100$ GHz from the central frequency of the pump, leading to a total walk-off of 1.29 ns. An asymmetry is clearly visible with an enhanced fan observed in the trailing part of the shock due to the deceleration of the pump. As shown in Fig. 10(b), an opposite detuning would have the same effect on the leading edge. Fig. 10(c) displayed the full experimental mapping of this process and shows that the difference of velocities between the two waves tends to smooth the DSW in an asymmetric manner. For comparison, we have also depicted the trajectory of the pump pulse in the timeframe of the CW landscape (dashed-line) which exactly follows the dispersive walk-off imposed by the frequency detuning between the two waves. This walk-off roughly achieves half of the pump width for $\delta f = 100$ GHz (displayed in yellow in Fig. 10(c)). Indeed, the shock waves naturally vanish when the difference of velocities between the two waves becomes larger than the speed of expansion of the shock – dictated by the maximum of frequency chirp imposed on the CW landscape (here around 70 GHz). This behavior is fully reproduced by numerical simulations, shown in Fig. 10(d) and is qualitatively consistent with the impact of cross-phase modulation as already reported in the context of short pulse generation [39].

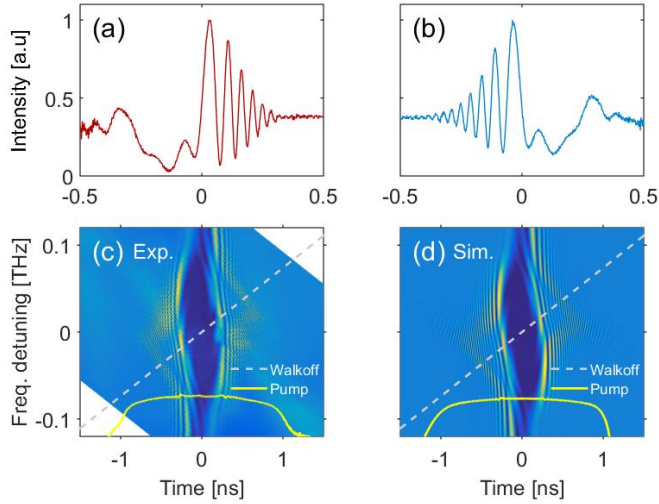

Fig. 10. Impact of the frequency detuning between probe and pump waves on the resulting temporal intensity profile of the probe for a pump power of 2 W. (a) Results obtained for an offset frequency of
$\delta \mathcal{F}$ = 100 GHz (red curve) and (b) $\delta \mathcal{F}$ = –100 GHz (blue curve). (c) and (d): Evolution of the output intensity profile of the probe wave as a function of frequency detuning $\delta \mathcal{F}$. Experimental results (c) are compared to numerical simulations in (d). Data are normalized by the input level of the CW landscape and are linearly mapped onto the full color range. Moreover, the dashed line corresponds to the walk-off between the two waves as a function of the frequency detuning while the yellow curve illustrates the output intensity profile of the pump in a Log scale.

## 5. IMPACT OF RANDOMNESS IN THE PROBE LANDSCAPE

We finally experimentally investigate the impact of randomness on the shock wave imposed by a partial incoherence in the probe landscape. Some examples of the impact of randomness in coupled fiber optics systems have already been experimentally studied in the past, but the coupling mechanism was essentially driven by a gain process [40, 41]. On the contrary, let us recall that our configuration does not imply any transfer of energy. To generate the partially coherent probe we spectrally filter an incoherent source, i.e. the amplified spontaneous emission from an Erbium doped fiber amplifier. The optical bandpass filter used has a Gaussian shape, is centered at the same frequency as the pump pulse, and its full width at half maximum $\Delta F$ can be tuned between 10 and 40 GHz, leading to temporal fluctuations of the intensity profile ranging from 44 to 11 ps. The filtered wave is then polarized orthogonally to the pump wave using a polarizer and injected into the fiber with a 10-mW average power. In contrast to previous sections, for which the shock wave was detected by means of a sampling oscilloscope, here the temporal output intensity profile is monitored with a 50-GHz real-time oscilloscope so as to resolve the randomness of the landscape [42].

In Fig, 11(a), we show one thousand distinct oscilloscope traces corresponding to different realizations of a ballistic shock wave formation on an input incoherent landscape with 10 GHz spectral width. When the pump is switched on, one can clearly see the typical holes emerging from the fluctuations of the 10 GHz partially coherent wave (panel (a1)). We can also make out the existence of bumps localized on each side of the temporal gap, typical of the ballistic shock. The details of the fan structure are however completely blurred. Moreover, details of the deterministic pump are reported in panel (a2) and show that the level of fluctuations may strongly vary along the pulse width. Indeed, significant intensity fluctuations surrounding the central part of the pulse can be observed, which are attributed to the XPM induced by the incoherent probe wave around the rarefaction area. In contrast, the central part and the wings of the parabolic profile remain fully coherent.

In order to further assess the statistical behavior of the waves, we have also evaluated the probability distribution function (PDF) of the intensity profiles along different parts of the waves. Results are summarized in panels (c1) and (c2) for the probe and pump, respectively. Whereas the rarefaction area of the probe (open blue circles, blue region in panel (a)) presents a narrow distribution centered close to zero, the bumps (red filled circles, red regions in panel (a)) are characterized by events presenting a higher intensity than the one appearing without any interaction with the pump (yellow crosses, yellow region in panel (a)). Regarding the pump, we observe that the PDF of the central part is much narrower than the parts affected by fluctuations.

These results are strongly affected by the degree of incoherence of the probe. Indeed, results in panel (b) reveal that increasing further the amount of disorder tends to hamper the shock wave until its total inhibition for $\Delta F$ higher than 40 GHz. We attribute this behavior to the fact that increasing the disorder into the probe landscape enhances the diffusion of the medium through chromatic dispersion, thus invalidating the shallow water approximation. Indeed, the temporal random fluctuations in the probe profile, typically much faster than the pulse duration, are highly diffusing and thus hamper the shock formation. These results can be also easily understood by analogy with a disordered material or a discrete medium for which randomness-induced scattering will absorb the shock, just like previously observed in an aqueous dye solution of silica spheres pumped by a Gaussian laser beam [12].

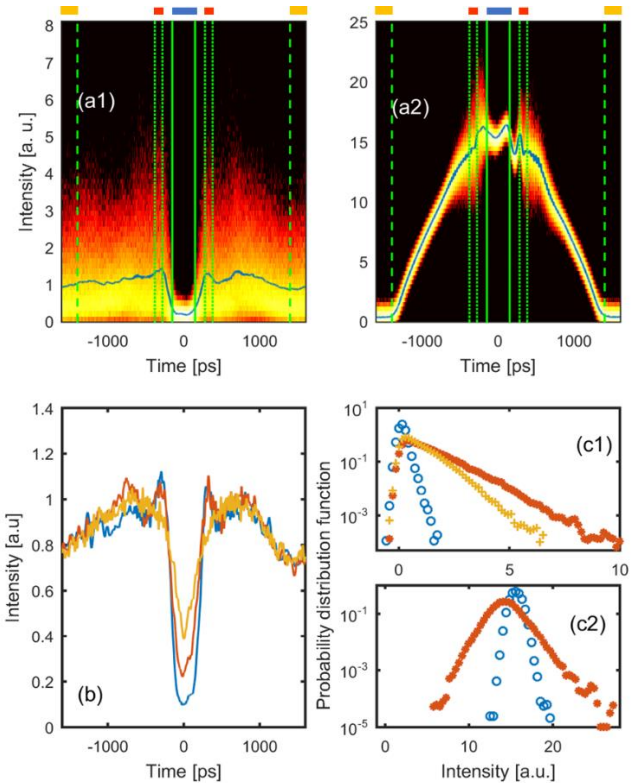

Fig. 11. Impact of partial coherence of the probe signal. Experimental results obtained for a pump peak power of 3.8 W. (a) Output landscape for a spectral coherence of 10 GHz. Results for the probe and pump are plotted on panels (a1) and (a2) respectively. (b) Influence of the coherence of the probe: average profile of the probe obtained for a coherence $\Delta F$ of 10, 20, 40 GHz are plotted with blue, red, and yellow lines respectively. (c) Evolution of the PDF of the various parts of the probe and pump waves (panel c1 and c2 respectively). Statistics for the central part of the probe (blue part in panels (a)) are plotted with blue open circles whereas properties in the lateral bumps (red part in panels (a)) are plotted with red full circles and with yellow crosses for the region outside the temporal pump extend (yellow part in panels (a)). Intensities are normalized with respect with the average intensity of the probe.

In order to confirm our experimental results, we have carried out a set of numerical simulations based on coupled equations (1). The initial incoherent wave can be modelled in the spectral domain by a Gaussian intensity profile of FWHM width $\Delta F$ with a delta-correlated random spectral phase uniformly distributed between $-\pi$ and $\pi$. Results after propagation in the fiber are averaged over 1000 shots (Fig. 12), which enables us to reproduce the PDF with sufficient accuracy. We confirm that in the case of low incoherence, the main features of the spectro-temporal structure of the ballistic shock wave as shown in Fig. 2 for the coherent case are qualitatively reproduced (see panels a). This is not the case anymore when the spectral width of the partially coherent width is of the same order as the frequency shifts that are induced through cross phase modulation and the central rarefaction zone tends to vanish, in agreement with our experimental observations. Using numerical simulations, we also tested the case of a 100 GHz initial incoherence for which the bandwidth of the electrical detection was not sufficient (note that optical approaches such as time microscopy could efficiently be experimentally implemented to accurately characterize the PDF of those incoherent waves [43]). For a 100 GHz incoherence, the ballistic shock wave has been fully blurred.

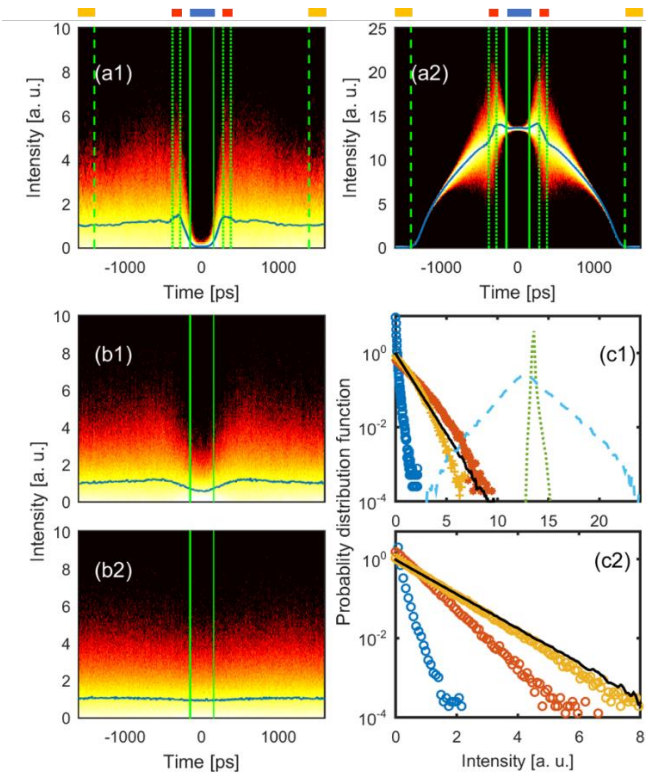

Fig. 12. Impact of the partial coherence of the probe signal. Results obtained from numerical simulations of the temporal intensity profile. (a) Output landscape for a spectral coherence of 10 GHz. Results for the probe and pump are plotted on panels (a1) and (a2) respectively. (b) Influence of the coherence of the probe: results obtained for a coherence of 20 and 100 GHz are plotted on panels (b1) and (b2) respectively. The blue lines in panels (a) and (b) are for the average behavior. (c) Evolution of the PDF of the various parts of the waves. Panel (c1) summarize the properties for a coherence of 10 GHz. PDF of the central part of the pump (located between the continuous green lines in panel a2) is plotted with a green dotted line whereas properties in the lateral fluctuating zone (between dotted lines in panels (a)) is plotted with a blue dashed line. Properties of the probe are plotted with circles for different regions: in yellow, for the region outside the temporal pump expansion (outside the dashed zone plotted in panels (a)), in red for the region corresponding to the bump (between dotted green lines of panel (a)), and in blue for the central region. Panel (c2): statistical properties of the probe in the central region for different levels of incoherence: results obtained for coherences of 10 GHz, 40 GHz and 100 GHz are reported with blue, red and yellow circles respectively. In panels (c), the continuous black line represents the input probe properties (corrected from the linear losses). Results are averaged over 1000 shots.

The statistical distribution of the temporal intensity profile (panel c1) is also in qualitative agreement with the trends reported experimentally. In the region of the humps, as the dispersion translates the XPM-induced incoherent phase into power fluctuations, the pump exhibits a broad distribution of the peak-power (with a ratio of the mean value $M$ by standard deviation $\sigma$ by the of $M/\sigma = 6$). On the contrary, the pump is unaffected close to its maximum ($M/\sigma > 100$). The PDF of the probe is strongly depleted in the central part, highlighting the nearly complete depletion this area even in presence of an initial incoherent landscape. The part of the pulse that does not interact with the pump (yellow circles) is less tailed than the initial distribution (characterized by a linearly decreasing PDF when plotted on logarithmic scale [44]), in agreement with the evolution in the normal regime of dispersion reported for turbulent waves [42]. On the contrary, the PDF evaluated by the hump regions exhibits higher values than the exponential distribution of the input Gaussian probe field.

Finally, the PDF calculated for the central region of the probe at different levels of initial incoherence (panel c2) confirms that increasing incoherence progressively close the gap opened by the ballistic shock so that the PDF gets closer and closer from the input PDF.

## 6. CONCLUSIONS

In conclusion, we report the first experimental demonstration of ballistic dispersive shock waves in the context of nonlinear optics. The principle is based on the nonlinear XPM interaction between a weak CW probe co-propagating in a normally dispersive optical fiber with an orthogonally polarized intense short pulse. In contrast to previous studies for which the scalar approximation has for the moment focused most of the attention, here the present shock wave is generated in a vectorial configuration. This cross-polarized interaction allows us to clearly mimic superfluid DSWs characterized by the appearance of an expanding temporal depletion of the central part surrounded by two repulsive oscillating fronts resting on the initial continuous landscape. The temporal expansion of the shock fans as well as its oscillating behavior can be well predicted by the maximum frequency chirp induced by the pump on the probe wave and the spectro-temporal mapping induced by chromatic dispersion. We have also experimentally investigated the impact of a frequency detuning leading to a difference of velocities between the two waves, and found that a large asymmetry and smoothing of the shock may be induced. Finally, we have studied the impact of randomness on the CW landscape, and conclude that a partially coherent wave, for which strong and fast temporal fluctuations drive a diffusion process, are able to hamper or even fully inhibit the shock dynamics.

All the experimental results are in full agreement with numerical simulations based on a Manakov system of coupled nonlinear Schrödinger equations. We anticipate that the presented results will stimulate analytical and statistical research that is above the scope of the present paper. We have shown how the pump drives the evolution towards the shock and continuously affects its evolution. Future works may for example focus on how the use of a chirped pump pulse could be beneficial to control the longitudinal evolution of the probe wave and associated DSWs.

More generally, shock wave phenomena are often difficult to study in their original environment, and also hard to reproduce in laboratory. Therefore, we believe that the present numerical and experimental research fully demonstrates that fiber optics may constitute a remarkable test-bed platform for the characterization of DSWs. Moreover, in contrast to numerous environments, here fiber components constitute a new avenue to produce and study ballistic dispersive shock waves in a non-destructive manner, by means of a compact, highly stable and high repetitive system. We therefore expect that other fields of physics investigating two component systems may be inspired by this research [45].

**Funding Information.** Institut Universitaire de France (IUF), Labex ACTION program (contract ANR-11-LABX-01-0001). Julien Fatome acknowledges the financial support from the European Research Council (Grant Agreement 306633, PETAL project) as well as the Conseil Régional de Bourgogne Franche-Comté (International Mobility Program) which has allowed him to visit The University of Auckland to contribute to this work as well as Agence Nationale de la Recherche, ANR program APOFIS. Miro Erkintalo acknowledges support from the Rutherford Discovery Fellowships administered by the Royal Society of New Zealand.

**Acknowledgment**. All the experiments were performed on the PICASSO platform in ICB. We also thank S. Trillo, M. Conforti and D. Castelló-Lurbe for fruitful discussions.